\documentclass{amsart}
\usepackage{amssymb}
\usepackage{amsmath}
\usepackage{graphicx} 
\usepackage{color}
\usepackage{xcolor}
\usepackage{appendix}
\usepackage[sep=5pt, offset=1em]{simpler-wick}
\usepackage{verbatim}
\usepackage{hyperref}

\theoremstyle{plain} 
\newtheorem{thm}{Theorem}[section]
\newtheorem{lem}[thm]{Lemma}

\theoremstyle{definition}

\newtheorem{expl}[thm]{Example}

\newtheorem*{defn*}{Definition}
\theoremstyle{remark}
\newtheorem{rem}[thm]{Remark}

\newcommand{\g}{\mathfrak{g}}
\newcommand{\mc}{\mathcal}
\newcommand{\mr}{\mathrm}
\newcommand{\bt}{\bullet}
\newcommand{\ra}{\rightarrow}
\newcommand{\M}{\mathcal{M}}
\newcommand{\wh}{\widehat}

\newcommand{\sfa}{\mathsf{a}}
\newcommand{\ad}{\mathrm{ad}}

\title[CS invariants from non-acyclic flat connections]{Perturbative Chern-Simons invariants from non-acyclic flat connections}
\author[P. Mnev]{Pavel Mnev}
\address{University of Notre Dame, Notre Dame, IN 46556, USA}
\address{Institut f\"ur Mathematik, Universit\"at Z\"urich, Winterthurerstrasse 190, CH-8057 Z\"urich,
Switzerland}
\email{pmnev@nd.edu}

\author[K. Wernli]{Konstantin Wernli}
\address{Centre for Quantum Mathematics, IMADA, University of Southern Denmark, Campusvej 55, 5230 Odense M, Denmark}
\email{kwernli@imada.sdu.dk}

\thanks{K. W. was supported by DFF Sapere Aude Grant 10.46540/4251-00015B}
\date{\today}

\begin{document}

\begin{abstract}
    We give a short review of our construction 
    of a higher-loop perturbative invariant of framed 3-manifolds, generalizing the perturbative Chern-Simons invariant of Witten-Axelrod-Singer, associated to an acyclic flat connection, to an invariant given by the integral of a certain ``Chern-Simons volume form'' over a smooth closed component of the moduli space of flat connections.
\end{abstract}

\maketitle

\section{Introduction}
In his pioneering work \cite{Witten89} Witten was studying the path integral of Chern-Simons theory
\begin{equation} \label{Chern-Simons path integral}
\int_{\mr{Conn}(\mc{P})/\mr{Gauge}}\mc{D}A\; e^{\frac{i}{\hbar}S_{CS}(A)} 
\end{equation}
-- an integral over the space of connections in a trivial $G$-bundle $\mc{P}$ over a 3-manifold $M$ modulo gauge transformations. 
Here 
\begin{equation}\label{S_CS}
S_{CS}(A)=\int_M \frac12\langle A \stackrel{\wedge}{,}dA\rangle + \frac16 \langle A \stackrel{\wedge}{,} [A\stackrel{\wedge}{,} A]\rangle
\end{equation}
is the Chern-Simons functional of the connection 1-form $A$.
Expression (\ref{Chern-Simons path integral}) is not defined as a measure-theoretic integral. However, Witten explained how to understand it, in the asymptotics $\hbar\ra 0$, as a stationary phase expansion. In particular, the leading and subleading order in $\hbar$ of a contribution of an \emph{acyclic}\footnote{A flat connection $A_0$ is said to be acyclic if de Rham differential twisted by $A_0$ has zero cohomology. In particular this implies that the gauge class of $A_0$ is an isolated point in the moduli space of flat connections.} flat connection $A_0$ were understood in \cite{Witten89}. Axelrod and Singer \cite{AS91}, \cite{AS94} constructed the higher order corrections to the stationary phase (``perturbative'') contribution of an acyclic flat connection to the path integral (\ref{Chern-Simons path integral}) as a formal power series in $\hbar$. Its coefficients are given as sums over graphs $\Gamma$ (``Feynman diagrams''), of 
integrals of certain differential forms $\omega_\Gamma$ over compactified configuration spaces of points on $M$. They proved that this series is an invariant of a framed 3-manifold, i.e., is independent of the choice of metric used to construct the forms $\omega_\Gamma$. 
Put differently, the metric determines the gauge-fixing for the Chern-Simons path integral and the result is independent of this choice.

In \cite{CSglob} we extended Witten-Axelrod-Singer construction to non-acyclic flat connections, constructing a volume form -- the ``Chern-Simons volume form''
-- on the ``nice'' (smooth and irreducible) stratum $\M'$ of the moduli space of flat connections $\M$. A change of Riemannian metric on $M$ is proven to change this volume form by an exact form, and hence integrals over closed components of $\M'$ yield invariants. 
The construction is based on building a family of effective actions on cohomology (``zero-modes''), parametrized by a background flat connection, and extracting a global volume form on the moduli space from it using the technology of formal geometry.

A part of motivation is the expectation that the integral of the Chern-Simons volume form over $\M'$ recovers the asymptotics $k\ra\infty$ of the Witten-Reshetikhin-Turaev invariant of the 3-manifold, under the assumption that $\M'$ is open dense in $\M$. More precisely, the WRT asymptotics is conjectured to be given by a sum over values of the Chern-Simons functional, and we expect that, if the preimage of such a value has a dense open subset of smooth irreducible flat connections, the corresponding contribution to the asymptotics coincides, up to appropriate normalization, with the integral of the Chern-Simons volume form. 

In the present note we give a short review of the main result of \cite{CSglob}:
we provide a concise definition of the invariant and sketch the proof of metric independence. In two appendices we briefly review the necessary elements of formal geometry, and its application to the moduli space of flat connections.

\section{The construction of the invariant}\label{sec: construction}
Let $M$ be a closed oriented smooth 3-manifold. Fix $G$ a compact simply-connected Lie group with Lie algebra $\g$. 
We fix some invariant nondegenerate bilinear form $\langle,\rangle$ on $\g$.\footnote{ 
\label{footnote: trace normalization}
E.g., the trace form associated to some representation.
For $G$ simple, it is conventional in the context of Chern-Simons theory to choose $\langle,\rangle$ to be $\frac{1}{h^\vee}$ times the Killing form, with $h^\vee$ the dual Coxeter number. In this case, $\frac{1}{8\pi^2}\langle x,x \rangle$ corresponds under Chern-Weil homomorphism to a generator of $H^4(BG,\mathbb{Z})$, which is related to invariance of $e^{\frac{ik}{2\pi}S_{CS}(A)}$ under large gauge transformations for $k\in\mathbb{Z}$,  cf. \cite{Witten89}, \cite{AS91}.
}
Also fix a Riemannian metric $g$ on $M$ (this is auxiliary data: the invariants we construct will turn out to be independent of $g$).

Given a flat connection $A$ in the trivial $G$-bundle $\mc{P}$ over $M$ (we identify connections with their 1-forms on $M$), we have the $A$-twisted Hodge decomposition of $\g$-valued forms on $M$ into $A$-harmonic forms (the kernel of $\Delta_A=d_A d^*_A+d^*_A d_A$ with $d_A=d+\mr{ad}_A$ the twisted de Rham differential and $d^*_A$ the twisted codifferential), $d_A$-exact forms and $d^*_A$-exact forms:
\begin{equation}\label{Hodge decomp}
\Omega^\bt(M,\g)=\mr{Harm}^\bt_A\oplus \mr{im}(d_A)\oplus \mr{im}(d^*_A).
\end{equation}
We denote:
\begin{itemize}
    \item $H^\bt_A$ the cohomology of the complex $(\Omega^\bt(M,\g),d_A)$.
    \item $i_A\colon H^\bt_A\ra \Omega^\bt(M,\g)$ the inclusion of cohomology as $A$-harmonic forms; $p_A\colon \Omega^\bt(M,\g)\ra H^\bt_A$ the projection onto cohomology via projecting onto harmonic forms.
    \item $P_A=i_A p_A\colon \Omega^\bt(M,\g)\ra \Omega^\bt(M,\g)$ the projector onto harmonic forms in (\ref{Hodge decomp}).
    \item $G_A=(\Delta_A+P_A)^{-1}\colon \Omega^\bt(M,\g)\ra \Omega^\bt(M,\g)$ the Green operator. 
    \item $K_A=d^*_A G_A\colon \Omega^\bt(M,\g)\ra \Omega^{\bt-1}(M,\g)$ the chain homotopy between identity and $P_A$. 
\end{itemize}

We say that a flat connection $A$ is irreducible if $H_A^0=0$ and is smooth if the $L_\infty$ algebra induced on $H_A^\bt$ from the dg Lie algebra $(\Omega^\bt(M,\g),d_A,[-,-])$ is trivial (has vanishing operations).\footnote{If $A$ is irreducible, then smoothness in this sense is tantamount to the moduli space being a smooth manifold in an open neighborhood of the gauge class of $A$.}
Let $\M$ be the moduli space of flat connections in $\mc{P}$ and let $\M'\subset \M$ be the stratum consisting of gauge classes of smooth irreducible flat connections.\footnote{Note that the tangent space to $\M'$ at $A$ is naturally identified with $H^1_A$.} 

Define
\begin{multline}\label{W}
    W= \Theta+\sum_\Gamma \frac{\epsilon^{l(\Gamma)-1}}{|\mr{Aut}(\Gamma)|}\Phi_\Gamma
    \quad \in C^\infty(\M', \mr{Sym}(H^1_A[1]\oplus H^2_A[-1])^*)[[\epsilon]]
    \\ 
    \cong \Omega^\bt(\M', \mr{Sym}(H^2_A[-1])^*)[[\epsilon]].
\end{multline}
Here $\epsilon$ is a formal infinitesimal parameter; $\Gamma$ runs over connected trivalent graphs with leaves (loose half-edges); $l(\Gamma)$ is the number of loops in $\Gamma$ (the first Betti number); $|\mr{Aut}(\Gamma)|$ is the order of the automorphism group of $\Gamma$. The weight of the graph is defined as
\begin{multline}\label{Phi_Gamma}
    \Phi_\Gamma= \\
    (-1)^\mu\int_{C_V(M)} \Big\langle  \bigwedge_{\mr{edges\;}e=(uv)} \pi_{uv}^*\wh\eta\; \wedge\; \bigwedge_{\mr{short\; loops\;}e=(vv)} \pi^*_v \wh\eta^\Delta \;\wedge \; \bigwedge_{\mr{leaves\;}l} \pi^*_{v(l)} \wh{i}(\sfa^2) , \bigotimes_{\mr{vertices}} f\Big\rangle_\Gamma.
\end{multline}
Here:
\begin{itemize}
    \item The integral is over Fulton-MacPherson-Axelrod-Singer compactification of the configuration space of $V$ distinct points on $M$ (with $V$ the number of vertices in $\Gamma$). Also, $\pi_{uv}\colon C_V(M)\ra C_2(M)$ is the map forgetting all points except $u$ and $v$. Similarly, $\pi_v\colon C_V(M)\ra M$ is the map forgetting all points except $v$. $v(l)$ in the third factor is the vertex incident to the leaf.
    \item The ``extended propagator'' $\wh\eta \in \Omega^2(\M'\times C_2(M),\g\otimes\g)[\epsilon]$ is, pointwise on $\M'$, the integral kernel of the operator
    \begin{multline}\label{Khat}
        \wh{K}=K - \epsilon K \mr{ad}^*_{i(\zeta)} G+ \epsilon^2 K \mr{ad}^*_{i(\zeta)} G \mr{ad}^*_{i(\zeta)} G  \\
        \in \bigoplus_{p=0}^2 \Omega^p(\M',\mr{End}(\Omega^\bt(M,\g))_{-1-p})[\epsilon]
    \end{multline}  
    with $\zeta=[\delta A]$ a placeholder for the tangent vector to $\M'$;\footnote{In other words, $\zeta=[\delta A]$ is the fiber coordinate on the odd tangent bundle $T[1]\M'$. 
    }
    $\mr{ad}_x^*\colon \omega\mapsto (-1)^{|\omega|} *\mr{ad}_x*\omega$ is the adjoint of the operator $\mr{ad}_x$ with respect to the Hodge inner product $\int_M \langle\alpha \stackrel{\wedge}{,} *\beta\rangle$. We are suppressing the subscript $A$ in $i,K,G$ for brevity.
    \item $\wh\eta^\Delta\in \Omega^2(\M'\times M,\g\otimes \g)[\epsilon]$ is the regularized evaluation of the integral kernel of $\wh{K}$ on the diagonal in $M\times M$.\footnote{The $(0,2)$-form component of $\wh\eta^\Delta$ on $\M'\times M$ is the term $L^{cont}$ in (PL5) in \cite{AS91}.}
    \item The ``extended inclusion'' $\wh{i}$ is
    \begin{multline}\label{ihat}
        \wh{i}=i - \epsilon G \mr{ad}^*_{i(\zeta)} i+ \epsilon^2 G \mr{ad}^*_{i(\zeta)} G \mr{ad}^*_{i(\zeta)} i \\
        \in \bigoplus_{p=0}^2 \Omega^p(\M',\mr{Hom}(H^\bt_A,\Omega^{\bt-p}(M,\g)))[\epsilon].
    \end{multline}  
    \item The zero-mode $\sfa^2$ is in $H^2_A[-1]$.
    \item $f\in \g^{\otimes 3}$ is the structure tensor of the Lie algebra $\g$.
    \item
    $\langle -,-\rangle_\Gamma$ is the extension of the 
    pairing
    on $\g$ to the pairing between $\g^{\otimes N}$ and $\g^{\otimes N}$ with $N$ the number of half-edges in $\Gamma$ 
    according to the incidence of half-edges to vertices in $\Gamma$. 
    Put another way, $\langle -,-\rangle_\Gamma=\langle \sigma_\Gamma(-),- \rangle$, where the pairing on the right matches $i$-th copy of $\g$ to $i$-th copy of $\g$ and $\sigma_\Gamma\in S_N$ is the permutation realizing the graph, see e.g. \cite[Section 3.10.2]{BVbook}.\footnote{
    One can think of a trivalent graph with leaves as a double coset
    $\Gamma\in {(S_V\ltimes S_3^V)\backslash S_N/((S_E\ltimes \mathbb{Z}_2^E)\times S_{\#\mr{leaves}})}$. In particular, a graph can be thought of as an equivalence class of permutations $\sigma_\Gamma$.
    }
    \item 
    $(-1)^\mu$ is a sign depending on the graph $\Gamma$: 
    \begin{equation}
        (-1)^\mu=(-1)^E (-1)^{\left[\frac{V}{2}\right]}(-1)^{\sigma_\Gamma}
    \end{equation}
    where the last factor is the sign of the permutation $\sigma_\Gamma$ above, $E$ is the number of edges.\footnote{
    A comment on the origin of the factors in the sign: $(-1)^E$ corresponds to the fact that if we are weighing the graph with $\epsilon^{-\chi(\Gamma)}$, with $\chi$ the Euler characterisitic, then the edge should be decorated with $-\wh\eta$ rather than $\wh\eta$ (or equivalently: vertices are decorated with $\frac{1}{\epsilon}f$ and edges with $-\epsilon \wh\eta$). $(-1)^{\sigma_\Gamma}$ is as in (3.10.2) in \cite{BVbook}. The middle factor is a Koszul sign arising from reshuffling of $V$ integrals over $M$ into a single integral over $M^{\times V}$ (which can be replaced -- by cutting out diagonals and compactifying -- by $C_V(M)$) in the free theory correlator (Gaussian average) $\langle \int_M \langle A, [A,A]\rangle \cdots \int_M \langle A, [A,A]\rangle \rangle_0$.
    }
\end{itemize}
Finally, $\Theta$ in (\ref{W}) is\footnote{One can think of $\Theta$ as the contribution of a special graph consisting of a single edge.} 
\begin{equation*}\label{Theta}
    \Theta=-\frac{1}{2\epsilon}\int_M \langle \wh{i}(\sfa^2) \stackrel{\wedge}{,} d \wh{i}(\sfa^2)\rangle = -\frac{\epsilon}{2} \langle \sfa^2, p\, \ad^*_{i(\zeta)}GdG \ad^*_{i(\zeta)} i(\sfa^2) \rangle,
\end{equation*}
where $d=d_A$. 

We define the ``Chern-Simons volume form'' on $\M'$ as the Berezin integral\footnote{ 
We use the following sign convention for the Berezinian  in (\ref{Y}): $\mc{D} \sfa^2 \,\mc{D}\zeta= \prod_i \mc{D} \sfa^2_i \,\mc{D}\zeta^i$ where $\zeta^i$ are coordinates on $H^1_A[1]$ associated to some basis $\xi_i$ in $H^1_A$ and $\sfa^2_i$ are coordinates on $H^2_A[-1]$ associated to the dual basis $\xi^i$ in $H^2_A$.
In our notations, for an odd line $\Pi \mathbb{R}$ parametrized by odd coordinate $\psi$, $\mc{D}\psi$ is the standard coordinate Berezinian satisfying $\int_{\Pi \mathbb{R}} \mc{D}\psi\;\psi=1$.
}
\begin{equation}\label{Y}
    Y= 
    \tau_A^{\frac12} \int_{H_A^2[-1]\oplus H_A^1[1]}
    \mc{D}\sfa^2\,\mc{D}\zeta
    \;e^{\langle  \zeta,\sfa^2 \rangle+W-W_0} \quad \in \Omega^{\mr{top}}(\M')[[\epsilon]].
\end{equation}
Here $W_0$ is the constant term in $\epsilon$ in $W$ and $\tau_A \in \mr{Det}H_A^\bt \cong (\wedge^\mr{top}(H_A^1)^*)^{\otimes 2}$ is the Ray-Singer torsion twisted by $A$ (and so $\tau_A^{\frac12}\in \Omega^\mr{top}(\M')$).

Next, we consider a framing on $M$ -- a trivialization of the tangent bundle $\phi\colon {M\times\mathbb{R}^3 \stackrel{\sim}{\ra} TM}$ (considered up to homotopy). 
Let 
\begin{equation*}
S_\mr{grav}(g,\phi)=\int_M \mr{tr} \left(\frac12 B\wedge dB + \frac16 B\wedge [B\stackrel{\wedge}{,}B]\right) \Big|_{B=\phi^*\nabla_{LC}}
\end{equation*}
be the Chern-Simons functional evaluated on the Levi-Civita connection associated to the metric $g$, trivialized using $\phi$. 
Here $\mr{tr}$ is normalized as minus half the trace of adjoint representation of $\mathfrak{so}(3)$.

\begin{thm}\label{main thm}
    There exists a universal formal power series\footnote{
    Coefficients of $c$ are (up to normalization) the anomalies $\beta_l$ appearing in \cite[Section 6]{AS94}.
    } in $\epsilon$ with coefficients depending only on the Lie algebra $\g$ (but not on $M$) 
    \begin{equation} \label{c(epsilon)}
        c(\epsilon)=-\frac{
        \langle f,f\rangle
        }{24 (2\pi)^2}\epsilon+O(\epsilon^3) \;\in \epsilon \mathbb{R}[[\epsilon^2]]
    \end{equation}
    such that, for any connected component $\M'_\alpha$ of $\M'$ which is a closed smooth manifold, the formal power series
    \begin{equation}\label{invariant}
        \Lambda=e^{c(\epsilon) S_\mr{grav}(g,\phi)}\int_{\M'_\alpha} Y\; \in \mathbb{R}[[\epsilon]]
    \end{equation}
    does not depend on the metric $g$ and thus is an invariant of the framed oriented 3-manifold $(M,\phi)$. 
    In (\ref{c(epsilon)}), $\langle f,f\rangle$ is the value of the $\Theta$-graph as a Jacobi graph for $\g$.\footnote{
    If one uses the pairing on $\g$ given by $\frac{1}{h^\vee}$ times the Killing form (cf. footnote \ref{footnote: trace normalization}), then $\langle f,f\rangle=-h^\vee \dim\g$.
    }
\end{thm}

\begin{rem}
    A change of orientation on $M$ acts on the invariant (\ref{invariant}) by $\epsilon\ra -\epsilon$.\footnote{
    Here is the idea: a flip of orientation on $M$ changes the weight of a graph by $\Phi_\Gamma$ by $(-1)^E (-1)^V=(-1)^{l(\Gamma)-1}$, where $(-1)^V$ come from the change of orientation on the configuration space $C_V(M)$ and $(-1)^E$ comes from the fact that each extended propagator $\wh\eta$ changes sign ($\wh{K}$ does not depend on orientation, but in transition to the integral kernel of an operator, the orientation is used). This $(-1)^{l(\Gamma)-1}$ can be absorbed into $\epsilon \ra -\epsilon$ in (\ref{W}). Additionally, Poincar\'e duality (which depends on the orientation of $M$) is used in the pairing $\langle \zeta,\sfa^2 \rangle$ in (\ref{Y}), so a flip of orientation can be absorbed into $\zeta\ra -\zeta$ or equivalently in $\epsilon \ra -\epsilon$ in the formulae for $\wh{K}$ and $\wh{i}$. 
    
    Another, ``physics,'' argument is: a change of orientation on $M$ acts on the Chern-Simons path integral $\int e^{\frac{i}{\hbar}S_{CS}(A)}$ by complex conjugation (since the action $S_{CS}$ is an integral of a 3-form on $M$ and flips sign with a change of orientation), assuming $\hbar$ real. Under the identification $\epsilon=-i\hbar$, cf. (\ref{Zglob vs Y}), this corresponds to $\epsilon\ra -\epsilon$.
    }
\end{rem}
\begin{rem}\label{rem: Zglob vs Y}
The global Chern-Simons partition function introduced in \cite{CSglob} is related to $Y$ by
\begin{equation}\label{Zglob vs Y}
    Z^\mr{glob}= e^{\frac{i}{\hbar}S_{CS}(A)}e^{\frac{\pi i}{4}\psi_A} Y\Big|_{\epsilon=-i\hbar}
\end{equation}
with 
$S_{CS}$
the Chern-Simons functional (\ref{S_CS}) and $\psi_A$ the Atiyah-Patodi-Singer eta-invariant of the operator $*d_A+d_A*$ acting on $\Omega^\mr{odd}(M,\g)$. Both $S_{CS}(A)$ and $\psi_A$ are locally constant functions on $\M'$.
\end{rem}

\begin{rem}
    In the case when $\M'_\alpha$ is an isolated point in the moduli space corresponding to an acyclic flat connection $A$, (\ref{invariant}) reduces (up to setting $\epsilon=-i\hbar$ and up to a prefactor $e^{\frac{i}{\hbar}S_{CS}(A)}e^{\frac{\pi i}{4}\psi_A+c_0S_\mr{grav}(g,\phi)}$, with $c_0=\frac{i\dim\g}{24(2\pi)}$) to the usual Witten-Axelrod-Singer invariant -- the perturbative path integral of Chern-Simons theory, over fluctuations around $A$.
\end{rem}

\begin{rem}
    Seifert fibered homology spheres provide an interesting class of 3-manifolds where $\M'$ has connected components that are closed smooth manifolds (as required in Theorem \ref{main thm}), 
    see  Theorem 5.1 in \cite{AHLMSS}.
\end{rem}

\section{The lowest (nontrivial) degree invariant}
\begin{figure}
    \centering
    \includegraphics[width=1\linewidth]{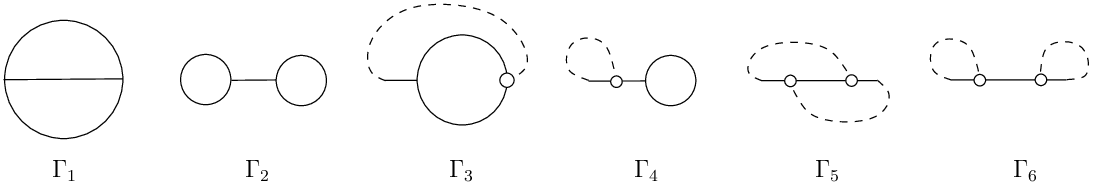}
    \caption{Six graphs contributing to (\ref{Lambda_1}). An edge/leaf with $n\in \{0,1,2\}$ circle vertices on it corresponds to the term of degree $n$ in $\zeta$ in $\wh{K}$, $\wh{i}$, or in $\Theta$
    (for graphs $\Gamma_5,\Gamma_6$). Dashed half-edges incident to circle vertices are decorated by $\zeta$ and solid leaves 
    are decorated by $\sfa^2$. Dashed edges show Wick contractions between $\zeta$ and $\sfa^2$ in (\ref{Y}).
    }
    \label{fig: six graphs}
\end{figure}

Expanding the invariant (\ref{invariant}) in powers of $\epsilon$ as $\Lambda=\Lambda_0 +\Lambda_1 \epsilon + \cdots$, we have
\begin{equation}\label{Lambda_0}
\Lambda_0=\int_{\M'_\alpha} \tau_A^{\frac12}
\end{equation}
and 
\begin{equation}\label{Lambda_1}
\Lambda_1=\int_{\M'_\alpha} \tau_A^{\frac12}\left(\Phi_1+\Phi_2+\Phi_3+\Phi_4+\Phi_5+\Phi_6- \frac{
\langle f,f \rangle
}{24(2\pi)^2}S_\mr{grav}(g,\phi)\right) 
\end{equation}
where
\begin{eqnarray*}
    \Phi_1 &=& 
    -\frac{1}{12} \int_{C_2(M)}f^{abc}f^{a'b'c'} \eta^{a a'}\wedge \eta^{b b'}\wedge \eta^{c c'}
    , \\
    \Phi_2 &=& 
    \frac18 \int_{C_2(M)} f^{abc}f^{a'b'c'} \pi_1^*(\eta^{\Delta})^{ab}\wedge \eta^{cc'}\wedge \pi_2^*(\eta^{\Delta})^{ a'b'}
    ,\\
    \Phi_3 &=& 
    -\frac12 \int_M f^{abc} ([K\ad^*_{\chi_i} G]^\Delta)^{ab}\wedge \chi^{ic}
    , \\
    \Phi_4 &=& 
    \frac12 \int_M f^{abc} (\eta^\Delta)^{ab}\wedge ( G\ad^*_{\chi_i}\chi^i)^c
    ,\\
    \Phi_5 &=& \frac12 \int_M
    \langle \chi^i\stackrel{\wedge}{,} \ad^*_{\chi_j} GdG \ad^*_{\chi_i}\chi^j \rangle, 
    \\
    \Phi_6 &=& -\frac12 \int_M
    \langle \chi^i \stackrel{\wedge}{,} \ad^*_{\chi_i} GdG \ad^*_{\chi_j}\chi^j \rangle.
\end{eqnarray*}
Indices $a,b,c,\ldots$ refer to an orthonormal basis in $\g$; 
summation over repeated indices is implied.\footnote{
Alternatively one can use an orthonormal basis with respect to minus the pairing on $\g$, which is convenient when the pairing is negative-definite. Then one should flip 
the signs of 
$\Phi_3,\Phi_4$.
}
For each point of $\M'_\alpha$ we choose arbitrarily a basis $\chi_i$ in $A$-harmonic 1-forms, and $\chi^i$ is the dual basis of harmonic 2-forms. The object $
[K\ad^*_{\chi_i} G]^\Delta
\in \Omega^1(M,\g\otimes\g)$ is the (regularized) evaluation of the integral kernel of the operator in brackets on the diagonal.\footnote{
One can also write $\Phi_3$ as an integral over the two-point configuration space 
$
\frac12 \int_{C_2(M)} f^{abc}\eta^{da}\wedge [\ad^*_{\chi_i}G]^{db}\wedge \pi_2^*\chi^{ic}
$, with $[\cdots]\in \Omega^2(C_2(M),\g\otimes\g)$ standing for the integral kernel of the operator in brackets. 
Yet another way to write $\Phi_3$ is as a regularized supertrace $-\frac12 \mr{Str}_{\Omega^\bt(M,\g)}^\mr{reg} \left(K\ad^*_{\chi_i} G \ad_{\chi^i} \right)$.
}
Expressions $\Phi_1,\ldots, \Phi_6$ correspond to the six graphs in Figure \ref{fig: six graphs}. Graphs $\Gamma_1,\Gamma_2$ are as in the usual Axelrod-Singer two-loop Chern-Simons invariant (at an acyclic flat connection), whereas the remaining graphs $\Gamma_3,\ldots, \Gamma_6$ account (in an appropriate sense) for the effect of globalization on the moduli space $\M'$. 

\begin{rem}
    Independence of $\Lambda_0$ on metric is immediate from independence of $\tau_A$ on the metric (a theorem of Ray and Singer \cite{RS}). In case of $\Lambda_1$, a variation of metric $g\ra g+\delta g$ induces a change of the integrand in (\ref{Lambda_1}) by an exact top-form $-d_{\M'}\Xi$ on $\M'_\alpha$ with 
    \begin{multline}\label{Xi}
        \Xi
        = 
        \iota_{\chi_i}\tau_A^{\frac12}\cdot
        \Big(
         \int_M \frac12 f^{abc} (\eta^\Delta)^{ab} \wedge (K\lambda_{\delta g}\chi^i)^c+
         \int_M \frac12 f^{abc} ([K\lambda_{\delta g} K]^\Delta)^{ab}\wedge \chi^{ic}\\
         + \int_M 
         \langle \chi^j\stackrel{\wedge}{,}\lambda_{\delta g} G \ad^*_{\chi_j}\chi^i \rangle
         - \int_M 
         \langle \chi^j\stackrel{\wedge}{,} \ad^*_{\chi_j}G \lambda_{\delta g} \chi^i \rangle
         \Big)
        \quad \in \Omega^{\mr{top}-1}(\M'_\alpha).
    \end{multline}
    with $\lambda_{\delta g}$ defined by (\ref{lambda}). 
    This expression for $\Xi$ arises as the degree $1$ term in $\epsilon$  in $\check{Y}^{\mr{glob},(1)}$, see (\ref{eq: metric invariance}).
    We refer the reader to Appendix \ref{Appendix: diagrammatic proof} for the sketch of direct diagrammatic proof of $\delta_g(\mr{integrand\;of\;}(\ref{Lambda_1}))=-d_{\M'}\Xi$.
    \begin{figure}
        \centering
        \includegraphics[width=0.8\linewidth]{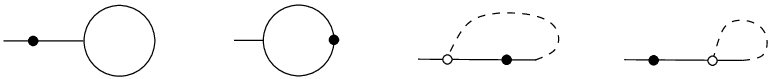}
        \caption{Four graphs contributing to (\ref{Xi}). The black vertex is decorated by $\lambda_{\delta g}$.}
    \end{figure}
\end{rem}

\section{Outline of proof of Theorem \ref{main thm}}
We give a schematic sketch of proof and refer to \cite{CSglob} for details. The proof relies on technology of Batalin-Vilkovisky formalism and formal geometry.

\textbf{Step 1:  constructing the extended effective action.} Let $\mr{FC}'$ be the space of smooth irreducible flat connections in $\mc{P}$ (not modulo gauge transformations), and let $\mr{Met}$ be the space of Riemannian metrics on $M$. One constructs the ``extended effective action on zero-modes $\sfa=\sfa^1+\sfa^2\in H^\bt_A[1]$'' -- a nonhomogeneous form
\begin{equation}\label{Wcheck}
    \check{W} = -\langle [\delta A],\sfa^2 \rangle+\frac{\epsilon}{2} \langle \sfa^2 ,\check\Theta \sfa^2 \rangle+
    \sum_\Gamma \frac{\epsilon^{l(\Gamma)-1}}{|\mr{Aut}(\Gamma)|}\check\Phi_\Gamma
    \quad \in \Omega^\bt(U,\wh{\mr{Sym}}(H_A^\bt[1])^*)[[\epsilon]]
\end{equation}
on an open neighborhood\footnote{The reason for restricting to a (sufficiently thin) open neighborhood of the diagonal in $\mr{FC}'\times \mr{FC'}$ is that the construction relies on ``desynchronized Hodge decomposition'' (discussed below) which is only guaranteed to exist for $A$ and $A'$ sufficiently close.
}
$U$ of $\mr{Diag}\times \mr{Met}$ in the space of triples $\mr{FC}'\times \mr{FC}'\times \mr{Met}=\{(A,A',g)\}$.\footnote{
A ``physics'' comment: the zero-form part of $\check{W}$ is the perturbative path integral of Chern-Simons theory over fluctuations around $A$, with Lorenz gauge-fixing $d^*_{A'}(\mr{fluctuation})=0$. Thus, $A',g$ are parameters of gauge-fixing, whereas $A$ is the background field.
More precisely, one splits fluctuations into ``zero-modes'' $\sfa$ -- solutions of linearized equations of motion modulo linearized gauge transformations -- and the complement, and integrates out the complement.
1-form components of $\check{W}$ along $A',g$ correspond to the generators of canonical BV transformations (``homotopies'') incurred by shifts of $A',g$; higher-form components are -- in an appropriate sense -- higher homotopies. These statements are encoded in the horizontality equation (\ref{Ycheck horizontality}).
}
In (\ref{Wcheck}) the notations are as in (\ref{W}): $\Gamma$ ranges over connected trivalent graphs with leaves; weights of graphs are defined similarly to (\ref{Phi_Gamma}) with the propagator $\wh\eta$ replaced with 
$\check\eta \in \Omega^2(U\times C_2(M),\g\otimes \g)[\epsilon]$ -- the integral kernel
of the operator
\begin{equation}\label{Kcheck}
    \check{K}=
    K +\epsilon KHK+\epsilon^2 KHKHK 
    \in \bigoplus_{p=0}^2\Omega^p(U,\mr{End}(\Omega^\bt(M,\g))_{-1-p})[\epsilon]
\end{equation}
and $\wh{i}(\sfa^2)$ replaced with 
\begin{equation}\label{icheck}
\check{i}(\sfa)=i(\sfa) +\epsilon KH i(\sfa) +\epsilon^2 KHKH i(\sfa) + \epsilon K \delta A 
\in 
\Omega^{\bt+1}(U\times M,\g)[\epsilon].
\end{equation}
Also appearing in (\ref{Wcheck}) is 
\begin{equation}\label{Thetacheck}
    \check\Theta = -p HKH i \in \Omega^2(U,\mr{End}(H^\bt_A)_{-1}).
\end{equation}
To explain the notations, we first remark that,
given $(A,A',g)\in U$, one has the ``desynchronized Hodge decomposition'' associated to the pair of differentials $d_A, d^*_{A'}$:
\begin{equation}\label{desync Hodge}
    \Omega^\bt(M,\g)=\mr{Harm}_{A,A'}^\bt\oplus \mr{im}(d_A)\oplus \mr{im}(d^*_{A'})
\end{equation}
with the first term -- ``$(A,A')$-harmonic forms''  -- being $\mr{ker}(d_A)\cap \mr{ker}(d^*_{A'})$. In (\ref{Kcheck}), (\ref{icheck}), (\ref{Thetacheck}),
$i\colon H^\bt_A\ra \Omega^\bt(M,\g)$ is the embedding of cohomology as $(A,A')$-harmonic forms, $p\colon \Omega^\bt(M,\g)\ra H^\bt_A$ is the projection to cohomology via projecting to harmonic forms in (\ref{desync Hodge}),
and $K\colon \Omega^{\bt}(M,\g)\ra \Omega^{\bt-1}(M,\g)$ is the chain contraction associated to the decomposition (\ref{desync Hodge}). Finally, 
$H\in \Omega^1(U,\mr{End}(\Omega^\bt(M,\g)))$ is the 1-form of a natural connection $\nabla^\mr{Hodge}$ in the trivial bundle over $U$ with fiber $\Omega^\bt(M,\g)$, preserving the decomposition (\ref{desync Hodge}). Explicitly: $H=H_{\delta A}+H_{\delta A'}+H_{\delta g}$ with\footnote{In fact, $H_{\delta A}$ cancels out in (\ref{Kcheck}), (\ref{icheck}), (\ref{Thetacheck}).}
\begin{eqnarray}
    H_{\delta A} &=&  -\Big(K\ad_{\delta A} dK+K\ad_{\delta A} P + P\ad_{\delta A} K\Big), 
    \nonumber
    \\
    H_{\delta A'} &=& -\Big(dG\ad^*_{\delta A'}Kd + dG\ad^*_{\delta A'}P+P\ad^*_{\delta A'}Gd \Big), \nonumber\\
    H_{\delta g} &=& dK\lambda_{\delta g} Kd+  dK \lambda_{\delta g} P+ P\lambda_{\delta g} Kd. \nonumber
\end{eqnarray}
Here $d=d_A$, $P$ is the projector to $(A,A')$-harmonic forms, $G=([d_A,d^*_{A'}]+P)^{-1}$ is the Green's operator; 
\begin{equation}\label{lambda}
\lambda_{\delta g}=*^{-1}\delta_g * 
\end{equation} 
measures the variation of the Hodge star with metric.\footnote{
Formulae (\ref{Kcheck}), (\ref{icheck}), (\ref{Thetacheck}) can be seen as  originating from homological perturbation lemma. One starts with the chain contraction $(i,p,K)$ associated to desynchronized Hodge decomposition (\ref{desync Hodge}), promotes it to a ``family'' chain contraction from $\Omega^\bt(U,\Omega^\bt(M,\g)),d_A$ to $\Omega^\bt(U,H_A^\bt)$ with zero differential, and then deforms the differential upstairs $d_A\ra d_A+\delta_{A'}+\delta_g$. By homological perturbation lemma, one obtains the deformed chain contraction given by (\ref{Kcheck}), (\ref{icheck}) (without the $K\delta A$ term) and the deformed induced differential $\delta_{A'}+\delta_g+\check\Theta$, with $\check\Theta$ as in (\ref{Thetacheck}). Term $K\delta A$ in (\ref{icheck}) and term $\langle [\delta A],\sfa^2 \rangle$ in (\ref{Wcheck}) account for changing infinitesimally the point of expansion $A$ of the path integral.
}

Let $\mc{D}$ be the bundle over $U$ with fiber $\mr{Dens}^{\frac12,\mr{formal}}(H_A^\bt[1])[[\epsilon]]$ -- the space of formal half-densities on the space of zero-modes.
The key property of the extended effective action (\ref{Wcheck}) is that 
\begin{equation}\label{Ycheck}
\check{Y}=\tau_A^{\frac12} e^{c(\epsilon)S_\mr{grav}(g,\phi)}e^{\check{W}}
\in \Omega^\bt(U,\mc{D})[[\epsilon]]
\end{equation}
satisfies the horizontality equation
\begin{equation}\label{Ycheck horizontality}
    \mathbb{D} \check{Y} =0
\end{equation}
with the differential operator 
$$\mathbb{D}=\nabla^\mathbb{H}+\Delta_\sfa+\frac12\langle \sfa, F \sfa \rangle$$ 
the ``Gauss-Manin superconnection'' on $\mc{D}$. Here $\nabla^\mathbb{H}$ is the connection on the bundle $\mathbb{H}$ over $U$ with fiber $H^\bt_A$ induced from restricting $\nabla^\mr{Hodge}$ to harmonic forms (and then extended to a connection in $\mc{D}$ 
in a natural way -- via pushforwards of half-densities); $F\in \Omega^2(U,\mr{End}(\Omega^\bt(M,\g)))$ is its curvature; $\Delta_\sfa=\langle \frac{\partial}{\partial \sfa^1}, \frac{\partial}{\partial \sfa^2} \rangle$ is the canonical BV operator on the fiber of $\mc{D}$.

The proof of (\ref{Ycheck horizontality}) relies on evaluating $\sum_\Gamma \frac{\epsilon^{-\chi(\Gamma)}}{|\mr{Aut}(\Gamma)|} \int_{C_V(M)} d\omega_\Gamma$ (with $\omega_\Gamma$ the integrand in (\ref{Phi_Gamma}) with  $\wh\eta,\wh{i}(\sfa^2)$ replaced with $\check{\eta},\check{i}(\sfa)$; here $\Gamma$ is allowed to be disconnected) in two ways -- (1) using identities satisfied by $\check{K}, \check{i}(\sfa)$ and (2) using Stokes' theorem on the configuration spaces. In the latter, one has to account for the contributions of ``principal'' and ``hidden'' boundary strata of the configuration space (collapses of two points and collapses of $\geq 3$ points, respectively). 
For the contribution of the hidden strata, the analysis of \cite{AS94} goes through (in particular, only the full collapse of a connected component of a graph $\Gamma$ contributes). Principal strata contributions cancel out in the sum over graphs by Jacobi identity/IHX relation.
Comparing computations (1) and (2) yields (\ref{Ycheck horizontality}). Contributions of hidden strata to (\ref{Ycheck horizontality}) are compensated by the derivative along metrics of the $S_\mr{grav}$ term in (\ref{Ycheck}).

\textbf{Step 2: reduction to the moduli space.} 
 Let $\check{W}_0$ be the term in $\check{W}$ which is constant in $\epsilon$ and a 1-form along $U$. Then (\ref{Ycheck horizontality}) implies 
\begin{equation}e^{-\check{W}_0}\mathbb{D}e^{\check{W}_0} e^{-\check{W}_0}\check{Y} = (\mathbb{D} + \nabla^\mathbb{H}\check{W}_0 + \Delta_\sfa\check{W}_0 + \{\check{W}_0,\cdot\} + \frac12 \{\check{W}_0,\check{W}_0\})(e^{-\check{W}_0}\check{Y}) = 0. \label{eq: DY}\end{equation} Here $\{\cdot,\cdot\}$ denotes the BV bracket on $H^\bullet_{A}[1]$. 

Restricting to the diagonal $A = A'$ and expanding the operator in brackets, \eqref{eq: DY} becomes 
\begin{equation}
    (\nabla^\mathbb{H}_A + \delta_g + \Delta_\sfa + \Delta_\sfa\check{W}_0 + \{\check{W}_0,\cdot\} \\
    + \nabla^\mathbb{H}\check{W}_0 + \frac12\{\check{W}_0,\check{W}_0\} + \frac12\langle \sfa, F\sfa\rangle)e^{-\check{W}_0}\check{Y} =0.\label{eq: op}
\end{equation} 
By pulling back $e^{-\check{W}_0} \check{Y}$ to the diagonal $A'=A$, 
setting $\delta A'=\delta A$ to be harmonic and passing to the quotient by gauge transformations, one obtains
\begin{equation}
    \overline{Y}=\tau_A^{\frac12} e^{c(\epsilon)S_\mr{grav}(g,\phi)}e^{\check{W}-\check{W}_0}\Big|_{A'=A,\; \delta A'=\delta A\;\mr{harmonic}} 
    \in \Omega^\bt(\M'\times \mr{Met},\mc{D})[[\epsilon]]. \label{eq: }
\end{equation}
Also, denote $R = R_{\delta A} + R_{\delta g} \in \Omega^1(\M'\times \mr{Met},\mc{D})$ the image of $\check{W}_0$. 
One then obtains from  \eqref{eq: DY} the equation
\begin{equation}
    (
    \nabla^G + \delta_g
    +\Delta_\sfa+\alpha) \overline{Y}=0.
\end{equation}
Here the first term in the brackets is the Grothendieck connection $\nabla^G = \nabla^\mathbb{H} + \{R_{\delta A},\cdot\} + \Delta_\sfa R_{\delta A}$ associated to the natural sum-over-trees formal exponential map\footnote{See Appendix \ref{app: formal geometry} for a definition of formal exponential maps and their associated Grothendieck connections, and Appendix \ref{app: sum over trees} for the definition of the sum-over-trees exponential map on $\M'$.} on $\M'$ (Lemma \ref{lem formula grothendieck}). The 1-form $\alpha$ is obtained by restriction of $\{R_{\delta g},\cdot\} + \Delta_\sfa R_{\delta g}$, making the expression in brackets a flat superconnection.\footnote{In fact, a restriction of the Grothendieck connection on $\M'\times \mr{Met}$, plus $\Delta_\sfa$ term, see Lemma \ref{lem: grothendieck M' times Met}. } Finally, the remaining terms in \eqref{eq: op} cancel when passing to the moduli space.\footnote{There is a 
subtlety in this proof having to do with the restriction to harmonic tangent vectors (and the fact that the notion of a harmonic form changes when one moves on the moduli space), for which we refer to \cite[Lemma 5.11]{CSglob}.} 

\textbf{Step 3: constructing the global object on the moduli space.} 
One has a chain map of complexes
\begin{equation}\label{p}
    (\Omega^\bt(\M'\times \mr{Met},\mc{D}), 
    \nabla^G+\delta_g
    +\Delta_\sfa+\alpha)  \xrightarrow{\mathsf{p}} (\mr{Dens}^{\frac12}(T^*[-1]\M')\wh\otimes \Omega^\bt(\mr{Met}) ,\delta_g+ \Delta_{\M'})
\end{equation}
It is constructed by homological perturbation lemma, deforming the differential of the complex on the left from $\nabla^G$.\footnote{This is a general fact about formal exponential maps depending on a parameter, see Appendix \ref{sec: parameters}.}
The map $\mathsf{p}$ sends the cocycle $\overline{Y}$ of the complex on the left  to a cocycle 
\begin{equation}\label{Yglob}
\check{Y}^\mr{glob}=\mathsf{p}\overline{Y}
\end{equation}
of the complex on the right. 

Expanding in form degrees along $\mr{Met}$, we have
$\check{Y}^\mr{glob}=\check{Y}^{\mr{glob},(0)}+\check{Y}^{\mr{glob},(1)}+\cdots$ with $\check{Y}^{\mr{glob},(0)}=e^{c(\epsilon) S_\mr{grav}(g,\phi)}Y$ -- the integrand in (\ref{invariant}).\footnote{Formula (\ref{Y}) for $Y$ was obtained in \cite{CSglob} from the computation of the r.h.s. of (\ref{Yglob}) where the key ingredient is the explicit formula for the projection  (\ref{p}) coming from homological perturbation lemma:
$\mathsf{p}(\cdots)=\Big( e^{\langle \frac{\partial}{\partial\sfa^2},\frac{\partial}{\partial [\delta A]} \rangle} (\cdots)\Big)\Big|_{\sfa^1=0,\, [\delta A]=0}$. The exponential of the differential operator in the r.h.s. can also be written as the Berezin integral (\ref{Y}).}
We exploit the fact 
that one can use the odd Fourier transform to identify half-densities on $T^*[-1]\M'$ equipped with BV differential $\Delta_{\M'}$ with forms on $\M'$ equipped with de Rham differential $d_{\M'}$, see (\ref{half-densities vs forms}) below.\footnote{See \cite{Witten90}, \cite{Schwarz}, \cite{Severa} for details.}

The cocycle condition for $\check{Y}^\mr{glob}$ restricted to form degree 1 along $\mr{Met}$ yields
\begin{equation}
    \delta_g \check{Y}^{\mr{glob},(0)} = - d_{\M'} \check{Y}^{\mr{glob},(1)}. \label{eq: metric invariance}
\end{equation}
Thus, the cohomology class of $\check{Y}^{\mr{glob},(0)}$ on $\M'$ is invariant under changes of the metric, which proves Theorem \ref{main thm}.
\section{Discussion}
We have 
constructed an extension of the perturbative Chern-Simons invariants to smooth irreducible components of the moduli space. To construct our invariant, we needed to introduce new diagrams beyond the previously known Chern-Simons Feynman diagrams. 

One could ask if there are other (potentially simpler) ways to generalize the perturbative Chern-Simons invariants to non-acyclic flat connections. One thing one can do is: fix a non-acyclic flat connection $A$ and a metric $g$, and consider the ``BV effective action on zero modes'' -- the form degree zero part of $\check{W}$ introduced in equation 
\eqref{Wcheck}, evaluated at $(A,A,g)$. One can show (\cite{CSremarks},\cite{CSlectures}) that it is independent of the metric, up to a $\Delta_\sfa$-exact term, where $\Delta_\sfa$ denotes the BV Laplacian on formal half-densities, and thus defines an invariant. However, if $\dim H^1_A >0$, then the relevant BV cohomology, and therefore also this invariant, is trivial.\footnote{The effective action has ghost number 0, and so corresponds (under the odd Fourier transform) to a top form. The cohomology of formal differential forms however, by Poincar\'e Lemma, is concentrated in degree 0 (corresponding to ghost degree $-\dim H^1_A$). } 
\footnote{A side remark: In \cite[Proposition 6]{CSremarks} one has a situation -- the effective BV action of Chern-Simons theory on a rational homology sphere computed at zero connection -- where the BV cohomology is trivial, but one can define a certain refinement of BV cohomology, quotienting BV cocycles by a restricted class of coboundaries coming from changes of gauge-fixing in the path integral. This refinement is nontrivial, and the refined BV cohomology class of the effective action is a meaningful invariant. However, 
it is not clear
how to define such a refinement for irreducible smooth connections.
}

Another way would be to consider the top form on $\M'$ given by taking  the zero form component of $\overline{Y}$, setting the fiber coordinates to zero and interpreting the result as a density/top form on $\M'$. This amounts to taking only diagrams appearing in the Axelrod-Singer construction, considering them in a family over the reference flat connection $A$, and multiplying them with the square root of the Ray-Singer torsion. In the lowest degree, this means considering only diagrams $\Gamma_1$ and $\Gamma_2$ in Figure \ref{fig: six graphs}.  At this order, one can check \eqref{eq: metric invariance} explicitly by a diagrammatic computation, see Appendix \ref{Appendix: diagrammatic proof}. In this computation, one term in the metric variation of $\Phi_{\Gamma_1} + \Phi_{\Gamma_2}$ is not cancelled by the de Rham differential of the 1-loop terms. 
The remaining term, however, is cancelled in the metric variation of $\Phi_{\Gamma_3}$, cf. cancellation $(1,3)+(3,4)+(7,6)=0$ in (\ref{diagrammatic proof: cancellations}).
This suggests that the variation of the top-form including only the Axelrod-Singer diagrams with respect to the metric is not exact, and thus its cohomology class does not define an invariant. Only after summing over all diagrams in Figure \ref{fig: six graphs} does one obtain a metric invariance up to exact terms. 
One reason these corrections arise is the fact that to obtain a global, equivariant family of chain contractions $(i_A,p_A,K_A)$, we need to vary the ``gauge-fixing operator'' $d_{A}^*$ together with the flat connection $A$ we are expanding around. 
\appendix
\section{Formal Geometry on the moduli space of flat connections}
In an attempt at self-containedness, we review certain aspects of formal geometry on the moduli space of flat connections. Formal geometry has been used extensively in the context of perturbative BV quantization, see \cite{CF}, \cite{BCM}. 
On the moduli space of flat connections it was considered in \cite[Section 2]{CSglob}.
\subsection{Elements of formal geometry} \label{app: formal geometry}
We briefly review some elements of formal geometry.

\subsubsection{Formal exponential maps} 
Let $X$ be a manifold. A \emph{generalized exponential map on $X$} is a map $\varphi\colon TX \supset U \to X, (x,y) \mapsto \varphi_x(y)$ defined on a neighborhood of the zero section $U \subset TX$ satisfying for all $x \in X$ that $\varphi_x(0) = x$ and $(d\varphi_x)_0 = \mathrm{id}_{T_xX}$. We consider two generalized exponential maps equivalent if all their vertical jets agree for every $x \in X$ and call the equivalence classes \emph{formal exponential maps}. 
A formal exponential map has a well-defined Taylor expansion map $T\varphi^* \colon C^\infty(X) \to \Gamma(X,\wh{\mr{Sym}}\ T^*X)$, defined as follows. For a function $f \in C^\infty(X)$, the pullback $\varphi^*f \in C^\infty(U)$ defines, for every $x \in X$, by restriction, a smooth function $f_x$ on  the open neighborhood $U_x = U \cap T_xX$ of $ 0 \in T_xX$. By Taylor expanding $f_x$ around $0 \in T_xX$ we obtain an element $T(\varphi^*f)_x \in \wh{\mr{Sym}}\ T^*_xX$.
\subsubsection{Grothendieck connection}
For a generalized exponential map, we  define the Grothendieck connection $\nabla^G$ of $\varphi$  -- a fiber bundle connection on an open neighborhood of the zero section in $TX$ -- by its parallel transport: For $x,x'\in X$ sufficiently close, the parallel transport of $y \in T_xX$ to $x' \in X$ is given by $y' = \varphi_{x'}^{-1}(\varphi_x(y))$. Since the parallel transport does not depend on the path from $x$ to $x'$, the Grothendieck connection is flat.  For a vector $\xi \in T_xX$ 
one then has a differential operator $\nabla_\xi^G\colon C^\infty(U) \to C^\infty(U)$ defined by taking a derivative of the parallel transport of $\nabla^G$ along a curve with tangent vector $\xi$. By taking Taylor expansion at the origin, one obtains a flat vector bundle connection on the vector bundle $\widehat{\mr{Sym}}\ T^*X$.\footnote{This is the image of the Cartan connection on the infinite jet bundle of the trivial line bundle $J^\infty\underline{\mathbb{R}}$ under the isomorphism $J^\infty\underline{\mathbb{R}} \cong \widehat{\mr{Sym}}\ T^*M$ induced by $\varphi$.} Being a flat vector bundle connection it extends to a differential, also denoted $\nabla^G$, on the complex $\Omega^\bullet(X, \widehat{\mr{Sym}}\ T^*X)$. An important fact is that this complex resolves functions on $X$. In fact, the Taylor expansion map $T\varphi^* \colon C^\infty(X) \to \Omega^0(X, \wh{\mr{Sym}}\ T^*X)$ is a quasi-isomorphism with quasi-inverse $p$ given by sending a form to its degree zero component and evaluation on the zero section. 
\subsubsection{Formal differential forms}
 The discussion above extends to other natural bundles over $X$, i.e., bundles derived from the tangent bundle. In the case of differential forms, one has the bigraded vector space $\Omega^\bullet(X,\wh{\mr{Sym}}\ T^*X \otimes \wedge^\bullet T^*X)$. This bigraded vector space has two natural differentials: $\nabla^G$ with bidegree $(1,0)$ and $d_y$, the fiberwise de Rham differential, which has bidegree $(0,1)$. The cohomology of $\nabla^G$ is concentrated in bidegree $(0,\bullet)$ and given by $T\varphi^*\Omega^\bullet(X)$, whereas the cohomology of $d_y$ is concentrated in bidegree $(\bullet,0)$ and given by $\Omega^\bullet(X)$.\footnote{The cohomology of $d_y$ is fiberwise the de Rham cohomology of a formal neighborhood of $0$ in a vector space, which by Poincar\'e lemma is isomorhic to constants.}    
 Using homological perturbation theory, one can prove that the total complex $(\Omega^\bullet(X,\wh{\mr{Sym}}\ T^*X \otimes \wedge^\bullet T^*X), \nabla^G + d_y)$ is quasi-isomorphic to the de Rham complex of $X$, and that the Taylor expansion map $T\varphi^*\colon\Omega^\bullet(X) \to \Omega^\bullet(X,\wh{\mr{Sym}}\ T^*X \otimes \wedge^\bullet T^*X)$ is a quasi-isomorphism.\footnote{However, the quasi inverse of the Taylor expansion map is deformed, and now is non-trivial on all forms of total degree $k$ (as opposed to only forms of bidegree $(0,k)$).}
 \subsubsection{Formal half-densities}
 In our case it is natural to work with half-densities on $T^*[-1]X$. From now on we will assume that $X$ is oriented. Then
 one can prove that there is a canonical isomorphism 
 \begin{multline} \label{half-densities vs forms}
 \mr{Dens}^\frac12T^*[-1]X \cong C^\infty(T^*[-1]X) \otimes \Omega^{\dim X}(X) \\
 \cong  \Gamma(X,\wedge^{-\bullet}(TX)) \otimes \Omega^{\dim X}(X) \cong \Omega^{\dim X -\bullet}(X).
 \end{multline}
 Under this identification, the canonical BV Laplacian $\Delta_X$ on half-densities corresponds to the de Rham differential on $X$ \cite{AlbertoBVLaplacian}. In particular, the results of the previous paragraph hold for the bicomplex  $(\Omega^\bullet(X,\mathcal{D}),\nabla^G, \Delta_y)$, where $\mathcal{D}$ is the bundle of formal half-densities on $T^*[-1]X$,\footnote{In this appendix we drop the dependence on the formal parameter $\epsilon$ as opposed to the main text.} considered as a bundle over $X$,
 $$ \mathcal{D}
 = \wh{\mr{Sym}}\ (T^*X \oplus T[1]X) \otimes {\det}^{\frac12}(T^*X \oplus T[1]X),
 $$
 with $\Delta_y$ the fiberwise BV Laplacian on formal half-densities. The cohomology of $\nabla^G$ is given by Taylor expansions of global half-densities, while the total complex is quasi-isomorphic to the complex of global half-densities $\mr{Dens}^\frac12(T^*[-1]X),\Delta_X)$.
\subsubsection{Including parameters}\label{sec: parameters} We now want to consider the situation where the formal exponential map $\varphi$ depends on an additional parameter $t$, running over an open subset $T \subset V$ of a vector space $V$. We are interested in  the complex $\Omega^\bullet(T \times X, \mathcal{D})$ -- forms on $T \times X$ with values in formal half-densities on $X$. On it, there are three different differentials: The de Rham differential $d_T$ along $T$, the Grothendieck connection $\nabla^G$, and the vertical BV operator $\Delta_y$, however, their sum no longer defines a differential since $d_T$ and $\nabla^G$ do not commute (because $\varphi$ depends on $t$). However, 
we can consider the formal exponential map on $T\times X$ given by $(x,t,y,\tau)\mapsto (\varphi_t(x,y),t + \tau)$ and its Grothendieck connection $\nabla^{G,T\times X}$.
Restricting the Grothendieck connection to formal half-densities which are constant along the formal $T$ directions, i.e. sections of the pullback bundle\footnote{Notice that this subbundle is preserved by the Grothendieck connection.} $\pi^*_X\mathcal{D}$ with $\pi_X \colon T \times X \to X$ the projection, we obtain a differential on $\Omega^\bullet(T\times X, \pi_X^*\mathcal{D})$. Since this bundle is trivial along the $T$ direction, we can write $\nabla^{G, T\times X}|_{\pi_X^*\mathcal{D}} = \nabla^{G,X} + d_T + L_\mu $ with $\mu \in \Omega^{1,0}(T \times X, \wh{\mr{Sym}}\ T^*X \otimes TX)$ a one-form along $T$ with values in formal vertical vector fields along $X$ and $L_\mu$ denotes Lie derivative with respect to $\mu$. Flatness of $\nabla^{G,T\times X}$ then implies 
\begin{equation}
\begin{aligned}
    d_{T}\mu + \frac{1}{2}[\mu, \mu] = 0 \\
    \nabla^G \mu + [\nabla^G,d_{T}] = 0
    \end{aligned}
\end{equation}
so that the operator of total degree one on $\Omega^\bullet(T\times X, \pi_X^*\mathcal{D})$ given by 
\begin{equation}
   D = d_T + \nabla^G + L_\mu  + \Delta_y = \nabla^G + \kappa
\end{equation}
squares to 0. By homological perturbation theory, 
considering $\kappa$ as a perturbation of $\nabla^G$, one obtains a quasi-isomorphism \begin{equation} p\colon (\Omega^\bullet(T\times X, \pi_X^*\mathcal{D}), D) \to (\Omega^\bullet(T,\mr{Dens}^\frac12(T^*[-1]X)), d_{T} + \Delta_X).
\end{equation}
\subsection{Formal geometry on $\mathcal{M}'$. } \label{app: formal geometry M prime}
We now explain how picking a metric on $M$, we obtain a formal exponential map on the smooth irreducible part $\M' \subset \M$ of the moduli space of flat $G$-connections on $M$. This brings us in the setup of the previous subsection, where $T = \mr{Met}$ is the space of metrics and $X = \M'$. 
\subsubsection{Sum-over-trees formal exponential map}\label{app: sum over trees}
Let $A$ be a flat connection on $M$ and $\sfa \in H^1$, and $g$ a Riemannian metric on $M$. In the notation of Section \ref{sec: construction}, we have a chain contraction (defined by the metric) $(i_A, p_A, K_A)$ of the complex $\Omega^\bullet(M,\g), d_A$ onto its cohomology $H^\bullet(M,\g)$. The chain contraction provides us with an explicit realization of the $L_\infty$ operations $(l_n')_{n\geq 2} \colon \wedge^nH^\bullet(M,\g) \to H^\bullet(M,\g)$. Namely, we can express $(l_n)'$ as sum over binary rooted trees where leaves are decorated by $i_A$, vertices are decorated with $[\cdot,\cdot]$, internal edges are decorated with $K_A$ and the root is decorated with $p_A$.  We assume that $A$ is irreducible (i.e., $H^0_A = 0$)
and $A$ is smooth, i.e., $l_n'= 0$ for all $n$. One then has a map $\varphi^g_{A} \colon H^1_A \supset V \to \Omega^1(M,\g)$, defined by 
\begin{equation}
    \varphi^g_A(\sfa) = A + i_A(\sfa) - \frac{1}{2}K_A[\sfa,\sfa] + \ldots = A + \sum_{n=1}^\infty\sum_{T \in T_n} \frac{(-1)^{n-1}}{|\mr{Aut}(T)|}\mu_T(\sfa,\ldots,\sfa) \label{eq: sum trees}
\end{equation}
where $T_n$ denotes isomorphism classes of binary rooted trees with $n$ leaves (we allow for the tree consisting of a single leaf), $\mr{Aut}(T)$ are the automorphisms of a $T$, and $\mu_T$ is obtained in the same way as $l_n'$, except that we now place $K_A$ at the root.\footnote{Under the assumption that $l_2' = \ldots = l_{n-1}' = 0$, the representative of $l_n'$ (given by evaluating the trees without the root) is closed. Vanishing of $l_n'$ then means this representative is exact, and summing $\mu_T$ over trees with $n$ leaves we obtain a particular primitive.} In \cite{CSglob} we show that there is a neighborhood $V \subset H^1_A$ on which \eqref{eq: sum trees} converges and that $\varphi^g_A(\sfa)$ is again flat.\footnote{The map $\varphi^g$ can be defined by \eqref{eq: sum trees} on a neighborhood of $0 \in H^1_A$ for all flat connections, but in general $\varphi_A(\sfa)$ is flat iff $\sfa \in MC(H^\bullet_A,(l_n)')$ is a Maurer-Cartan element of the induced $L_\infty$ operations.} Notice that $(i_A,p_A,K_A)$ is equivariant with respect to the action of the gauge group, hence the same is true for $\varphi^g_A$. Therefore, $\varphi^g_A$ descends to an open neighborhood of the zero section of $T\M'$ (which is the quotient of the cohomology bundle $\mathbb{H}$ over $\mr{FC}'$) that we denote by $\underline{\varphi}^g$. The explicit formula \eqref{eq: sum trees} shows that $\underline{\varphi}^g$ defines a generalized exponential map on $\M'$. 
\subsubsection{A formula for the Grothendieck connection of $\underline{\varphi}^g$}
Let $\check{W}_0 $ be the 1-form part of the evaluation of \eqref{Wcheck} at $\epsilon = 0$, explicitly $\check{W}_0$ is given by 
\begin{equation}
    \check{W}_0 = -\langle [\delta A], \sfa^2\rangle + \sum_T\frac{\epsilon^{-1}}{|\mr{Aut} (T)|}\check{\Phi}_T \in \Omega^1(U,\wh{\mr{Sym}}(H^\bullet_A[1])^*)
\end{equation}
where $T$ runs over binary trees. Notice that $\check{W}_0$ is linear in $\sfa^2$, i.e. we have $\check{W}_0 \in \Omega^1(U,\wh{\mr{Sym}}(H^1_A)^*\otimes (H^2_A)^*)$. 
Restricting $\check{W}_0$ to the diagonal, restricting to harmonic tangent vectors and passing to the quotient, we obtain a 1-form
\begin{equation}
    R = R_{\delta A} + R_{\delta g} \in \Omega^1(\M' \times \mr{Met}, \widehat{\mr{Sym}}\ T^*\M' \otimes T\M')
\end{equation}
with $R_{\delta A}$ and $R_{\delta g}$ the 1-form components along $\M'$ and $\mr{Met}$ respectively. We denote $R_{\delta A,g}$ the restriction of $R_{\delta A}$ to the slice of fixed $g$ in $\M' \times \mr{Met}$. Then we have the following result:
\begin{lem}[\cite{CSglob}]\label{lem formula grothendieck}
    The Grothendieck connection $\nabla^{G,g}$, induced by the formal exponential map $\underline{\varphi}^g$, on the bundle of formal functions on $\M'$ is given by 
    \begin{equation}
    \nabla^{G,g} = \nabla^{\mathbb{H}} + R_{\delta A,g}.
    \end{equation}
\end{lem}
In particular, on the bundle of formal half-densities, $\nabla^G$ acts as 
\begin{equation}
    \nabla^{G,g} = \nabla^\mathbb{H} +\{R_{\delta A,g},\cdot\} + \Delta_\sfa R_{\delta A,g}. 
\end{equation}

\subsubsection{Variation of the Grothendieck connection w.r.t metric} 
The formal exponential map $\underline{\varphi}^g$ depends on the Riemannian metric $g \in \mr{Met}$. To understand its dependence on $g$, one can, as in Appendix \ref{sec: parameters}, consider the formal exponential map on $\M' \times \mr{Met}$ defined by 
$\varphi([A],g)(\sfa,\sigma) \mapsto (\underline{\varphi}^g(\sfa),g+\sigma)$, for $\sigma \in S^2T^*M$ in a neighborhood of 0. Then, we have: 
\begin{lem}\label{lem: grothendieck M' times Met}
    The Grothendieck connection $\nabla^{G,\M' \times \mr{Met}}$ is given by 
    \begin{equation}
        \nabla^{G,\M' \times \mr{Met}} = \delta_g + \nabla^{\mathbb{H}} + R_{\delta A} + R_{\delta g} - \delta g \frac{\delta}{\delta \sigma}.
    \end{equation}
\end{lem}
This Lemma was not stated in this form in  \cite{CSglob}, but its proof is analogous to the proof of Lemma \ref{lem formula grothendieck} by considering also variation along the metric. 
In particular, the restriction to the bundle of formal half-densities along $\M'$ is 
\begin{equation}
    D = \delta_g + \nabla^{\mathbb{H}} + \{R,\cdot\} + \Delta_\sfa R.
\end{equation}

\section{A sketch of direct diagrammatic proof of metric independence of $\Lambda_1$}
\label{Appendix: diagrammatic proof}
Metric independence of the invariant in degree $O(\epsilon^1)$, $\Lambda_1$ (\ref{Lambda_1}), is of course a part of Theorem \ref{main thm}, but one can also check it directly. We sketch the argument here (modulo signs), as a sanity check.

\begin{figure}
    \centering
    \includegraphics[width=1\linewidth]{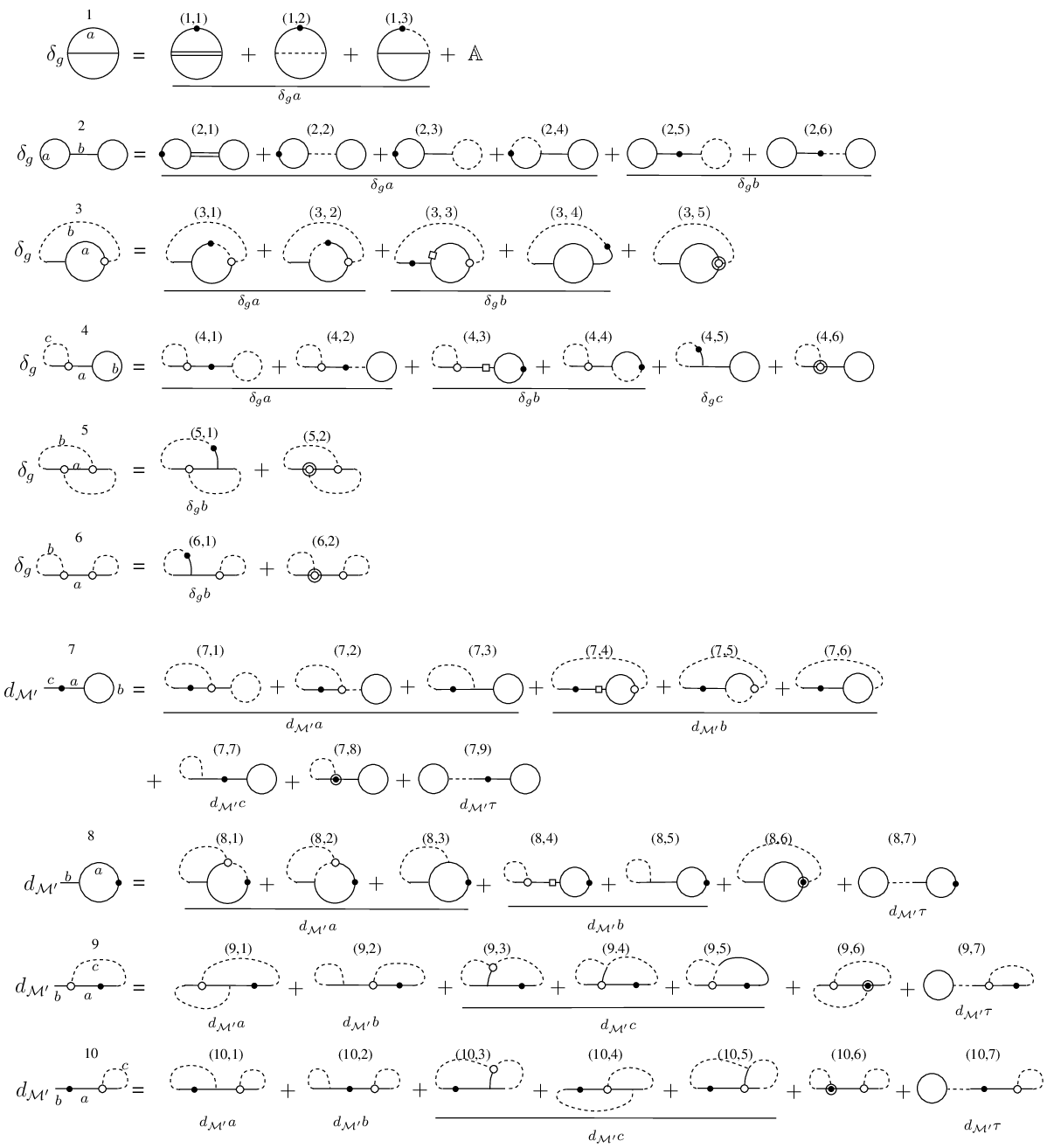}
    \caption{Diagrammatic formulae for $\delta_g$ applied to terms in $\lambda$ and $d_{\M'}$ applied to terms in $\Xi$.}
    \label{fig: diagrammatic proof}
\end{figure}

Denote 
\begin{equation}
\lambda=D_1+D_2+D_3+D_4+D_5+D_6+D_\mr{grav}
\end{equation}
the integrand in (\ref{Lambda_1}), with $D_i=\tau_A^{\frac12} \Phi_i$ for $i=1,\ldots,6$ and $D_\mr{grav}=-\frac{\langle f,f\rangle}{24(2\pi)^2} \tau_A^{\frac12}S_\mr{grav}(g,\phi)$. 
Also let $\Xi=D_7+D_8+D_9+D_{10}$ be as in (\ref{Xi}). (Recall that $\lambda$, $\Xi$ are the $O(\epsilon^1)$ terms in $\check{Y}^{\mr{glob},(0)}$ and $\check{Y}^{\mr{glob},(1)}$).
We want to prove that
\begin{equation}\label{delta_g lambda + d_M Xi=0}
    \delta_g \lambda+ d_{\M'}\Xi=0.
\end{equation}
One can do it directly by applying $\delta_g$ to terms in $\lambda$ and applying $d_{\M'}$ to terms in $\Xi$, write the results as sums of (weights of) certain Feynman diagrams -- see Figure \ref{fig: diagrammatic proof} -- and then track the cancellations of diagrams.

In Figure \ref{fig: diagrammatic proof}, for $i=1,\ldots,6$, we label the diagrams computing $\delta_g D_i$ by $(i,j)$. Likewise, for $i=7,\ldots,10$, we label the diagrams computing $d_{\M'} D_i$ by $(i,j)$.

This computation is based on identities
\begin{eqnarray*}
    \delta_g K &=& -[d, K H_{\delta g} K]-K H_{\delta g} P - P H_{\delta g} K,\\
    \delta_g i &=& -dK H_{\delta g} i,\\
    \mc{L}_\chi
    K &=& -[d, K H_{\delta A'=\chi} K]+K H_{\delta A'=\chi} P - P H_{\delta A'=\chi} K - K \ad_{\chi} K 
    ,\\
    \mc{L}_\chi i &=& -dK H_{\delta A'=\chi} i-K \ad_{\chi} i,\\
    \mc{L}_\chi \tau &=& \tau\cdot \mr{Str}\, K \ad_\chi.
\end{eqnarray*}
Here $\chi\in \mr{Harm}^1_A\cong T_A \M'$ and $\mc{L}_\chi$ is the directional derivative in the direction $\chi$ on the moduli space; when acting on the torsion $\tau$, $\mc{L}_\chi$ uses an appropriate connection in the determinant line bundle, see \cite[Lemma D.1]{CSglob}.\footnote{There are also some identities used for simplification, e.g., $K H_{\delta A'=d \beta} K=K \ad_\beta K$, $K H_{\delta A'=d\beta} P= K \ad_\beta P$, for $\beta\in \Omega^0(M,\g)$.}

In Figure \ref{fig: diagrammatic proof} we give labels $a,b,\ldots$ to elements of the Feynman graph  on the left (we pick one label for each orbit of $\mr{Aut}(\Gamma)$) and on the right we indicate which terms arise from the derivative hitting which element of the graph on the left.\footnote{Including the possibility that the derivative hits the torsion factor $\tau_A^{\frac12}$ implicitly contained in $D_i$. This happens in terms  $(7,9)$, $(8,7)$, $(9,7)$, $(10,7)$.}

The conventions in Figure \ref{fig: diagrammatic proof} (the assignment of tensors to edges and vertices -- the Feynman rules) are as follows.

\textbf{Edges:} A solid edge is assigned $K$; a dashed edge is assigned the projection onto harmonic forms $P$; a solid-dashed edge is the projection onto harmonic 2-forms if the solid end is treated as ``out'' and onto harmonic 1-forms if the solid end is ``in;'' a solid double edge is the identity;\footnote{
Identity on the double edge $e=(uv)$ is the interpretation if the diagram is seen as an integral over $M^{\times V}$. Alternatively, in terms 
of the 
compactified configuration space $C_V(M)$, one considers the integral over 
the principal boundary stratum where $u$ and $v$ approach one another (and other propagators connecting $u$ and $v$ are replaced with their continuous part at $u=v$.)
} a solid edge ending with a square box is $dK$.

\textbf{Vertices:} An ordinary trivalent vertex is the wedge product of forms tensored with the Lie bracket in $\g$; a white circle vertex is assigned $H_{\delta A'=\chi}$ with $\chi\in \mr{Harm}^1_A$ the argument coming from the side half-edge; a black vertex is $H_{\delta g}$. A circled black vertex is $\mathcal{L}_\chi H_{\delta g}$ 
($\chi$ decorates the side half-edge). A circled white vertex is $\delta_g H_{\delta A'=\chi}$.

Additionally, each diagram is understood to contain the torsion factor $\tau_A^{\frac12}$.

\textbf{Cancellations:}
The term $\mathbb{A}$ in $\delta_g D_1$ is the anomalous part of the metric dependence of the Theta graph, as in \cite{AS91}. It is cancelled by $\delta_g D_\mr{grav}$. Diagrams $(1,1)$ and $(2,1)$ cancel by Jacobi identity. 
The remaining diagrams cancel as follows:
\begin{multline}
\label{diagrammatic proof: cancellations}
    (1,2)+(8,3)=0;\;(1,3)+(3,4)+(7,6)=0;
    (2,2)+(8,7)=0;\\
    (2,3)+(8,5)=0;\;
    (2,4)+(4,5)+(7,3)=0;\;
    (2,5)+(7,7)=0;\;
    (2,6)+(7,9)=0;\\
    (3,1)+(8,1)=0;\;
    (3,2)+(10,4)=0;\;
    (3,3)+(7,4)=0;\;
    (3,5)+(8,6)+(5,1)+(9,3)=0;\\
    (4,1)+(10,2)=0;\;
    (4,2)+(10,7)=0;\;
    (4,3)+(8,4)=0;\;
    (4,4)+(6,1)+(10,1)=0;\\
    (4,6)+(7,8)=0;\;
    (5,2)+(9,6)=0;\;
    (6,2)+(10,6)=0;\;
    (7,1)+(9,2)=0;\\
    (7,2)+(9,7)=0;\;
    (7,5)+(9,1)+(10,3)=0;\;
    (8,2)+(9,5)=0;\;
    (9,4)+(10,5)=0.
\end{multline}
This proves (\ref{delta_g lambda + d_M Xi=0}).

\end{document}